# Non-invasive geophysical investigation and thermodynamic analysis of a palsa in Lapland, northwest Finland

Tomáš Kohout[1,2], Michał S. Bućko[1], Kai Rasmus[3,4], Matti Leppäranta[1], Ilkka Matero[1,5]

[1]Department of Physics, University of Helsinki, Helsinki, Finland (tomas.kohout@helsinki.fi)

[2]Institute of Geology, Academy of Sciences of the Czech Republic, Prague, Czech Republic

[3]Luode Consulting Oy, Jyväskylä

[4]SYKE, Freshwater Center, Jyväskylä

[5]Finnish Meteorological Institute, Helsinki, Finland

**Abstract**

Non-invasive geophysical prospecting and a thermodynamic model were used to examine the structure, depth and lateral extent of the frozen core of a palsa near Lake Peerajärvi, in northwest Finland. A simple thermodynamic model verified that the current climatic conditions in the study area allow sustainable palsa development.

A ground penetrating radar (GPR) survey of the palsa under both winter and summer conditions revealed its internal structure and the size of its frozen core. GPR imaging in summer detected the upper peat/core boundary, and imaging in winter detected a deep reflector that probably represents the lower core boundary. This indicates that only a combined summer and winter GPR survey completely reveals the lateral and vertical extent of the frozen core of the palsa. The core underlies the active layer at a depth of ~0.6 m and extends to about 4 m depth. Its lateral extent is ~15 m x ~30 m. The presence of the frozen core could also be traced as minima in surface temperature and ground conductivity measurements. These field methods and thermodynamic models can be utilized in studies of climate impact on Arctic wetlands.

**Introduction**

Palsas are distinctive features in sporadic permafrost regions of Arctic wetlands. The word 'palsa' (of Finnish or Saami (balsa) origin) denotes a hummock with an ice core rising out of a bog. Palsas contain a perennially frozen core of peat or till, and possess a life cycle in the order of 100 years (Seppälä, 1988, 2006; Pissart, 2002). Their growth and decay are driven by the surface heat balance with specific feedback mechanisms, with the critical driving factors being snowfall, snowdrift, liquid precipitation, and air temperature.

Knowledge of the growth and internal structure of palsas (e.g. Seppälä 1988,1995, Gurney 2001, Pissart 2002, Hofgaard 2003, Zuidhoff and Kolstrup, 2005, Seppälä 2006, 2011) and the extent of their frozen cores is limited to only a few investigations, where destructive methods were used. The palsa core usually consists of a perennially frozen peat or a till core with ice layers. The origin of the ice layers is discussed in the review by Seppälä (2011) and is either due to ice segregation, core buoyancy, or a combination of both. The core extends below the surrounding peat level and in a mature palsa stage can reach the mineral soil boundary. The core is covered by a peaty active layer. Modelling of the thermodynamics of palsas includes advanced numerical models (An and Allard 1995) and empirical climate models based on air temperature integrals (e.g., Fronzek et al., 2006).

We employ non-destructive geophysical methods to study the internal structure of a palsa and to map the lateral extent and depth of its frozen core. We combine both summer and winter survey data in order to study seasonal variations in the active layer state. To obtain a regional perspective, the birth and growth of palsas is approached by thermodynamic modelling using an analytic scaling model and forced by given climatic conditions. Together, these methods can be used to examine the physics of palsas in the permafrost zone and to evaluate the climate impact on Arctic wetlands.

**Study Site**

The studied palsa is located south of Kilpisjärvi village and close to Lake Peerajärvi, in northwest Finland (WGS84 68.884836º N, 21.051822º E; Figure 1). Several other palsas occur in the surrounding area. The palsa was selected because long time-series data of snow depth are available at the site (University of Helsinki, Department of Biosciences). The palsa rises ~3 m above the surrounding water-saturated peat. Although the southern slope of the palsa is well preserved, the northern part has been modified by adjacent road construction. Because the road base lies at the northern boundary of the palsa, our study focuses on its southern boundary. Cracking and erosion of the peat occur along the steep slopes of the palsa. Based on the classification in Seppälä (1988), Gurney (2001) and Hofgaard (2003), this palsa is a peat-cored, dome type, and probably in its early collapsing stage.

**Thermodynamics of Palsas**

The surface heat flux forces the growth and decay of the frozen core of palsas. Figure 2 shows the monthly mean air temperatures and snow depths in Kilpisjärvi. Because the annual mean air temperature is –2.0°C, the soil temperature at 10 m depth is probably near 0°C, and the heat flux at the lower boundary of the palsa is probably very small. The soil thermodynamics are then driven by surface fluxes and heat conduction, and the frozen core forms by phase changes after the freezing point has been reached.

Analytical models can be applied to a simple model of palsa birth (Leppäranta, 1993). In the absence of snow and ignoring the heat loss to deeper peat, starting with fully unfrozen peat, the thickness of the frozen layer after the first winter is

$$h_1 = \sqrt{\frac{2k_f}{\nu \rho_f L}(F - F_0)} \qquad (1)$$

where $\rho_f$ and $k_f$ are the density and thermal conductivity of the frozen peat, $\nu$ is the peat porosity, $L$ is the latent heat of freezing, $F$ is the sum of freezing-degree-days, and $F_0$ is the freezing-degree-days used for cooling the peat to 0°C. In the Kilpisjärvi region, on average $F$ = 1737°C day. Based on a heat budget analysis for Lake Kilpisjärvi region (Lei et al. 2012, Leppäranta et al. 2012), the first cold month (air temperature below 0°C) is used for the cooling, and then $F - F_0$ = 1726°C day. We can assume $\nu \approx 0.3$ at the site. The most critical factor is the thermal conductivity of peat; for $k_f \sim 1$ W m$^{-1}$ °C$^{-1}$ we have $h_1 \sim 1.8$ m. Adding snow reduces the growth of the frozen layer. Assuming that snow accumulation and the depth of the frozen layer are correlated (by climatology), $h_s \approx \lambda h_1$, Eq. (1) transforms into

$$h_1 = \sqrt{a(F-F_0)}, \quad a = \frac{2k_f}{(1+\lambda \frac{k_f}{k_s})\nu \rho_f L} \tag{2}$$

where $k_s$ is the conductivity of snow. For $k_f/k_s$ = 5, the thickness of the frozen layer is 1.5 m for $\lambda$ = 0.1 and 0.74 m for $\lambda$ = 1. The typical regional value of the parameter $\lambda$ is $\lambda \sim 0.7$ but it can be much smaller due to snowdrift. Since snowdrift is more pronounced from the top of a young palsa, heat fluxes at the peat surface are enhanced. Weaker insulation by snow on top of the palsa allows the ice core to grow faster after the initial formation.

Thawing of the frozen peat layer is a reverse process to that of growth. Snow cover must melt first. Then heat is conducted through the thawed peat to the surface of the frozen core to progress thawing as

$$\Delta h = \sqrt{\frac{2k_m}{\rho_m \nu L}(G - G_0)} \tag{3}$$

where $\rho_m$ and $k_m$ are the density and conductivity of thawed peat, $G$ is the sum of the positive degree-days, and $G_0$ stands for the positive degree-days used for snow melting. The positive degree-day method used for snow melting tells that the melting takes a time $G_0 = h_s/A$, where $h_s$ is the snow thickness and $A$ ~1 cm (°C day)$^{-1}$ is the degree-day coefficient. Kilpisjärvi climatology shows that on average $G$ = 1066°C day and $G_0$ = 100°C day. The condition for the initiation of perennial frozen ground, and palsa in peat land, is thus obtained from:

$$\frac{F-F_0}{G-G_0(h_s)} > \frac{\rho_f}{\rho_m} \cdot \frac{k_m}{k_f}\left(1+\lambda \frac{k_f}{k_s}\right) \tag{4}$$

For the Kilpisjärvi region, the left-hand side is > 1, while the right-hand side is within about 0.5–3.0, depending primarily on the net accumulation of snow (coefficient $\lambda$, by the approximation $h_s \approx \lambda h_1$). It is seen that with efficient wind-driven snow transport out of the site (small $\lambda$), the thickness of frozen peat is enough to persist over summer. In the case of no snow, the left hand side of (4) is 1.5 while the right hand side is 0.5 and the inequality is satisfied. But with normal snow protection of the ground ($\lambda$ ~0.7), we need $(F-F_0)/(G-G_0)$ > 2.3 which is unrealistically high. With $\lambda$ < 0.5, the inequality (4) would be satisfied. But for a finite thickness of the palsa core persist over summer, the parameter $\lambda$ needs to be much less than the limiting value of 0.5.

When a palsa has survived its first summer, its frozen core continues to thicken. The summer thawing is similar from year to year, but with less snow the thickness of the frozen soil can be doubled from the seasonal average. Spatial air temperature variations are rather smooth, based on comparison with nearby stations (see also Lei et al. 2012). In contrast, snow accumulation and snow drifting show large variability, which is reflected in the soil temperature evolution and therefore has the key role in the dynamics of palsas. In a growing palsa in the study region, the exchange of heat between the lower boundary of palsa and deeper peat is weak because of the small temperature differences. Degradation of palsas commences by active-layer erosion, that causes more summer melting. Thus, there is a decreasing trend of the active-layer thickness.

For a palsa to become thicker the long run, the heat loss at the bottom becomes significant. This growth takes place independently of the surface conditions:

$$\frac{dh_b}{dt} = -\frac{Q_b}{v\rho_m L} \tag{5a}$$

where $h_b$ is the depth of the palsa and $Q_b \sim -0.1$ W m$^{-2}$ is the heat flux at the bottom. This is of a magnitude of 3.5 cm year$^{-1}$. In addition, in cold years the thickness of the palsa may increase due to heat loss to the atmosphere:

$$h_{n+1} = \sqrt{h_n^2 + a(F - F_1)} \tag{5b}$$

where $h_n$ is the thickness of the palsa at the end of winter n, $a$ is a coefficient (see Eq. 2) and $F_1 < F$ are the freezing-degree-days used for freezing of the surface layer. Since the thawing takes place in the surface layer, there is no melting of the bottom of the palsa, but the depth increases monotonously until the degradation phase.

**Geophysical Survey Instruments and Methods**

Three geophysical survey methods were used to study the palsa: ground penetrating radar survey (GPR), frequency-domain electromagnetic ground conductivity mapping (EM), and thermal imaging.

GPR is a non-invasive, geophysical method that provides high-resolution 2D or 3D images of the subsurface (e.g. Davis and Annan 1989). GPR transmits radar wave pulses into ground, and the pulses are reflected from boundaries where a contrast in electric permittivity exists (e.g. frozen/unfrozen peat, ice/sediment). Thus, GPR is useful for mapping the near-surface structure of the partly frozen subsurface because of the strong dielectric permittivity contrasts between frozen and unfrozen materials (e.g. Doolittle et al. 1992, Horvath 1998, Moorman et al. 2003; Bradford et al. 2005, Ross et al. 2005, Brosten et al. 2006, Yoshikawa et al. 2006, Jørgenson and Andreasen 2007, Munroe et al. 2007, Kneisel et al. 2008). We used a Malå ProEx GPR with a 250 MHz shielded antenna to map the subsurface structure of the palsa, performing measurements in April and August 2011 along the same profile in a NE-SW direction. For comparison, a profile in a perpendicular NW-SE direction was measured in April 2013. The data were acquired continuously with 10 cm spacing (using a wheel as a trigger) while towing the GPR along the profile line. Stacking of between 2 and 32 traces (depending on the profile) was applied during the data acquisition. The radargrams were subsequently processed using RadEx software. DC and background removal, trace equalization, amplitude correction, migration, bandpass filtering, and topography correction procedures were applied.

The frequency-domain electromagnetic ground conductivity mapping was measured along the NE-SW profile in August 2011 with a Geonics EM-31 Mk. 2 electromagnetic sounding instrument working at 9.8 kHz with a 3.66 m coil spacing, in a way similar to Kneisel et al. (2008). Thermal infrared imaging variations in the surface temperature along the profile were measured in August 2011 on a cloudy day using a Control Company

xCite as: Kohout et al. 2014. Permafrost and Periglacial Processes, 25, 45–52. DOI: 10.1002/ppp.1798

Traceable Total-Range DT-380 infrared thermometer 1 m above the ground surface (spot size diameter 12.5 cm).

The surface topography of both the snow-covered (April) and the snow-free (August) palsa was measured using a Topcon AT-F4 theodolite. In April, the snow depth along the profile was measured using a manual snow probe. In August, a 3 m long core through the palsa was recovered using a manual soil probe (core diameter ~2 cm, length 1 m). Three 1 m long, cores were obtained using 1 m long extension bars. Unfortunately, the soil probe was lost during a subsequent sampling attempt (3-4 m depth).

**Results**

Thermodynamic Analysis

During winter 2010–2011 and summer 2011, the freezing-degree-days and positive-degree-days were $F = 1890°C$ day and $G = 1383°C$ day, respectively. In March 2011 the average thickness of snow was about 1 m in the Kilpisjärvi region according to the weather data of the Finnish Meteorological Institute. The snow depth on the top of the Peerajärvi palsa was 0.3 m (Figure 3), indicating that the winter growth was normal but summer thawing was likely greater than normal. Because the time-scale of palsa growth and collapse is several decades, a one year anomaly is not significant. Figure 4 compares the air temperatures in Enontekiö weather stations Näkkilä (374 m a.s.l.), Kilpisjärvi (480 m a.s.l.) and Saana (1004 m a.s.l.). In winter, a lower atmospheric temperature inversion results in the monthly mean air temperature increasing at higher elevations, whereas in summer the reverse occurs. Therefore variations in winter growth are small, but summer thawing depends on the altitude.

This also holds for long-term averages. At the Saana mountain weather station, the freezing-degree-days are similar to those at the Kilpisjärvi station (1737°C day), and so the inversion in atmosphere temperature is typical here in mid-winter. But the positive-degree-days are much less, on average 630°C day. Although there are no snow data for the top of the mountain, the snow cover there is normally thin and the permafrost condition (4) is well satisfied.

Field Survey

The winter and summer GPR profiles in the NE-SW direction are presented in Figures 5 and 6. Figure 3 shows the snow cover depth over the palsa. It is evident that the imaging capability of the radar differs in winter and summer conditions. In winter, the radar waves penetrate through the cold snow cover relatively easily into the soil. The areas with a deeper snow cover can be identified by the presence of the snow/peat reflector. One has to keep in mind that only structures comparable to or thicker than the radar signal wavelength can be detected.

The radar wavelength is a function of the wave velocity, which itself is a function of the dielectric permittivity of the medium through which the wave propagates. The permittivities of unfrozen peat, frozen peat, and snow are hard to determine as they depend on their exact composition, porosity, and liquid water content. Therefore the radar wave velocity was determined by comparing the apparent snow depth on the radargram to the snow depth measured using the snow probe. The best correlation was received for a radar signal velocity of ~10 cm/ns corresponding to a wavelength and approximate spatial resolution of ~40 cm. Thus in areas with a thin snow cover the snow/peat interface is invisible in radargrams.

A velocity of ~10 cm/ns is also expected to be similar in the frozen peat core (Moorman et al. 2003; Munroe et al. 2007) and was selected for the time to depth conversion and for the topography correction. Below the snow the winter radargram (Figure 5) is almost featureless up to approximately 80 ns two-way travel time (~4 m depth). This zone represents the frozen core of the palsa.

Below this zone a weak reflector is detected beneath the top of the palsa. We interpret this reflector as representing a lower boundary of the palsa's frozen core (Figure 5). The lateral extent of this reflector (and thus the width of the palsa core) is approximately 15 m. The upper boundary of the frozen core is not visible in the winter radargram. This is most likely due to the fact that in winter the overlying peat is also frozen, and thus there is no contrast in dielectric properties of the seasonally frozen active layer and perennially frozen peat core.

The summer radargram (Figure 6) was distinctly different to that of winter. First, the radar waves were significantly scattered as they reached the uneven and highly porous surface of the relatively dry peat active layer over the palsa. Below the surface, a strong shallow reflector is interpreted as a boundary between the unfrozen active layer and the frozen core of the palsa. The thickness of the palsa peat cover was determined from the recovered core to be 60 cm and correlated to a two-way travel time of approximately 25 ns in the radargram (Figure 6). Thus the velocity in the overlying, relatively dry active layer is ~5 cm/ns, which is lower compared to that in frozen peat. This value is similar to that reported by Doolittle et al. (1992) and Brosten et al. (2006). The velocity in the frozen core remains the same as in winter (~10 cm/ns). As the thickness of the frozen material is greater compared to that of active layer, the velocity of ~10 cm/ns was also selected for the topography correction. The 3 m long recovered core indicated that the frozen core, situated below the top peat cover, consisted of frozen peat with numerous ice lenses up to ~20 cm thick. The lower core boundary, visible in the winter radargram, was not identified in the summer radargram. This reflector was most likely hidden due to a significant portion of the radar energy being attenuated by the active layer in summer. A significant portion of the energy is lost at the top and bottom of the active layer.

Figure 7 presents a combined winter/summer radar profile with interpretations and time-to-depth conversion. The combined profile was created by superimposing the processed semi-transparent summer red-blue profile over the black-white winter one. For the winter data, a single-layer model with 10 cm/ns velocity was used for the time-to-depth conversion. For the summer data, a two-layer model is necessary due to the presence of the unfrozen active layer. Velocities of 5 cm/ns in the unfrozen active layer and 10 cm/ns in the frozen core are used. The stronger reflectors in the lefthand part of the profile are probably artefacts caused by the base of the road beside the palsa.

For comparison, Figure 8 shows a NW-SE winter profile completed in April 2013 in a perpendicular direction to that measured in 2011. The processing was similar to the 2011 profiles. The frozen core boundary can be traced at the bottom of the profile and the spatial length of the frozen core is at least 30 m in this direction. Possibly, the core continues beyond the profile towards the SE (Figure 8).

The surface temperature and electrical conductivity measurements in summer (Figure 9) revealed slightly lower temperature and conductivity values over the palsa.

**Discussion**

Comparison of the winter and summer GPR palsa surveys shows that it is not possible to accurately map the whole extent of the palsa using a single winter or summer radargram. Instead, the combined winter/summer radar profile is needed.

This may explain why the lower core boundary was not imaged during earlier GPR surveys performed in summer. Results of Doolittle et al. (1992) and Horvath (1998) show the active layer and underlying frozen palsa core, but not the reflector associated with the lower core boundary. As with our summer data, a significant amount of electromagnetic energy was reflected at the active layer / frozen core boundary and thus it was not possible to reliably observe deeper reflectors. A winter GPR survey may help in such cases because the active layer is frozen and has similar dielectric properties to the underlying frozen core. Thus, the electromagnetic signal reaching the frozen core is of higher energy.

Similarly to previous GPR palsa results, no significant reflectors are observed within the frozen core itself. This is probably because the ice layers often present within palsa cores are thin, and thus below GPR resolution, and have similar dielectric properties as the surrounding frozen peat. Unlike the profiles presented in Horvath (1998), no diagonal interfaces are identified in our data of the frozen core.

In the combined winter/summer profile, all of the main parts of the palsa (snow cover, peat cover, frozen core, unfrozen base material) are visible. It is also possible to determine the vertical thickness of the core to be approximately 3.4 m. Thus, both winter and summer data are needed to reconstruct the full extent of the frozen core.

The local surface temperature minima observed in summer result from the cooling effect of the palsa core. The frozen core and relatively dry overlying peat (active layer) also reduce the ground electric conductivity compared to surrounding wet peat, as observed in ground conductivity EM mapping.

Thermodynamic scaling analysis based on the analytic model presented here can estimate the thickness of seasonally frozen ground in the region and predict favourable conditions for palsa formation (e.g. thin snow conditions). The long-term growth of the palsa is possible due to heat losses to the deeper peat layers and to the atmosphere in cold years, when the surface layer becomes fully frozen before the end of winter. The relative importance of these two modes of growth cannot be solved from the present data. Sampling and deep temperature profiling are needed to better understand the long-term growth of palsas.

**Conclusions**

Only a combined winter and summer GPR survey of the palsa fully revealed its internal structure and the size of its frozen core. The frozen core extends vertically from ~0.6 m depth (as verified with the manual corer) to ~4 m depth, and its lateral extent along the profile was ~15 m and at least 30 m in the perpendicular direction. Winter snow depth measurements and coring results were used to calibrate the radar wave velocity that is essential in time-to-depth conversion of the radargrams. The coring also indicated ice lenses (up to ~20 cm thick) within the frozen core, a size below the GPR resolution (~40 cm). The presence of the frozen core can also be traced from the surface temperature and ground conductivity data as local minima.

The favourable palsa-forming conditions were confirmed by thermodynamic modelling. The analytic model is an effective tool for interpreting the influence of climate changes on palsa zones.

Non-destructive geophysical methods, especially the GPR survey, proved to be suitable for palsa studies and can be potentially applied in future studies of a larger number of palsas. The dual winter/summer measurement and topographical correction is essential to identify the extent of the frozen core of the palsa.

**Acknowledgments**

This work has been supported by the Academy of Finland project no. 140939. Institute of Geology, Academy of Sciences of the Czech Republic is supported by Ministry of Education, Youth and Sports project no. RVO67985831. Authors would like to thank to Friederike Gehrmann for language corrections.

**References**

An, W. and Allard, M. 1995. A mathematical approach to modelling palsa formation: Insights on processes and growth conditions. *Cold Regions Science and Technology* **23**, 213-244.

Brosten, T.R., Bradford, J.H., McNamara, J.P., Zarnetske, J.P., Gooseff, M.N., Bowden, W.B. 2006. Profiles of Temporal Thaw Depths beneath Two Arctic Stream Types using Ground-penetrating Radar. *Permafrost and Periglacial Processes* **17**, 341–355 doi: 10.1002/ppp.566

Bradford, J.H., McNamara, J.P., Bowden, W., Gooseff, M.N. 2005. Measuring thaw depth beneath peat-lined arctic streams using ground-penetrating radar. *Hydrological Processes* **19**: 2689–2699 doi:10.1002/hyp.5781

Davis, J. L. and Annan, A.P. 1989. Ground-penetrating radar for high-resolution mapping of soil and rock stratigraphy. *Geophysical Prospecting* **37**: 531-551.

Doolittle, J.A., Hardisky, M.A., Black, S. 1992. Ground-Penetrating Radar Study of Goodream Palsas, Newfoundland, Canada. *Arctic and Alpine Research* **24**, 173-178.

Fronzek, S., Luoto, M. and Carter, T.R. 2006. Potential effect of climate change on the distribution of palsa mires in subarctic Fennoscandia. *Climate Change* **32**, 1-12.

Gurney, S. D. 2001. Aspects of the genesis, geomorphology and terminology of palsas: perennial cryogenic mounds. *Progress in Physical Geography* **25** (2): 249–260. doi: 10.1177/030913330102500205

Hofgaard, A. 2003. Effects of climate change on the distribution and developments of palsa peatlands: background and suggestions for a national monitoring project. *Norwegian Institute for Nature Research project report 21*, 33 pages.

Horvath, CL. 1998. An evaluation of ground penetrating radar for investigation of palsa evolution, Macmillian Pass, NWT, Canada. In *Permafrost: Seventh International*

*Conference Proceedings*, Lewkowicz, A.G., Allard, M. (eds). Centre d'e´tudes Nordiques, Universite´ Laval, Sainte-Foy. Collection Nordicana **57**, 473–478.

Jørgensen, A.S., Andreasen, F. 2007. Mapping of permafrost surface using ground-penetrating radar at Kangerlussuaq airport, western Greeland. *Cold Regions Science and Technology* **48**: 64–72 doi: 10.1016/j.coldregions.2006.10.007

Kneisel, C., Hauck, C., Fortier, R., Moorman, B. 2008. Advances in Geophysical Methods for Permafrost Investigations. *Permafrost and Periglacial Processes* **19**, 157–178 doi: 10.1002/ppp.616

Lei, R., Leppäranta, M., Cheng, B., Heil, P. and Li, Z. (2012) Changes in ice-season characteristics of a European Arctic lake from 1964 to 2008. *Climatic Change* **115**: 725–739. DOI 10.1007/s10584-012-0489-2

Leppäranta, M. 1993. A review of analytical sea ice growth models. *Atmosphere–Ocean* **31**(1): 123–138.

Leppäranta, M., Shirawawa, K. and Takatsuka, T. 2012. Ice season in Lake Kilpisjärvi in Arctic tundra. *Proceedings of the 21st IAHR Ice Symposium*.

Moorman, B. J., Robinson, S.D., Burgess, M.M. 2003. Imaging periglacial conditions with ground-penetrating radar. *Permafrost Periglacial Processes* **14**: 319–329 doi: 10.1002/ppp.463

Munroe, J.S., Doolittle, J.A., Kanevskiy, M.Z., Hinkel, K.M., Nelson, F.E., Jones, B.M., Shur, Y., Kimble, J.M. 2007. Application of Ground-Penetrating Radar Imagery for Three-Dimensional Visualisation of Near-Surface Structures in Ice-Rich Permafrost, Barrow, Alaska. *Permafrost and Periglacial Processes* **18**: 309–321 doi: 10.1002/ppp.594

Pissart, A. 2002. Palsas, lithalsas and remnants of these periglacial mounds. A progress report. *Progress in Physical Geography* 26: 605-621. doi: 10.1191/0309133302pp354ra

Ross, N., Harris, C., Christiansen, H.H., Brabham, P.J. 2005. Ground penetrating radar

investigations of open system pingos, Adventdalen, Svalbard. *Norsk Geografisk Tidsskrift - Norwegian Journal of Geography* **59**: 129-138. doi: 10.1080/00291950510020600

Seppälä, M. 2011. Synthesis of studies of palsa formation underlining the importance of local environmental and physical characteristics. *Quaternary Research* **75**, 266-370 doi:10.1016/j.yqres.2010.09.007

Seppälä, M. 2006. Palsa mires in Finland. *The Finnish environment* **23**: 155–162.

Seppälä, M., 1995. How to make a palsa: a field experiment on permafrost formation. *Zeitscrift für Geomorphologie N.F. Supplement-Band* **99**: 91–96.

Seppälä, M. 1988. Palsas and related forms. In Advances in periglacial geomorphology, Clark, M.J. (ed.). Wiley: Chichester; 247–278.

Seppälä, M. 1982. An experimental study of the formation of palsas. *Proceedings Fourth Canadian Permafrost Conference, Calgary*. National Research Council of Canada: Ottawa; 36–42.

Yoshikawa, K., Leuschen, C., Ikeda, A., Harada, K., Gogineni, P., Hoekstra, P., Hinzman, L., Sawada, Y., Matsuoka N. 2006. Comparison of geophysical investigations for detection of massive ground ice (pingo ice). *Journal of Geophysical Research* **111**: E06S19. doi:10.1029/2005JE002573

Zuidhoff, F.S., Kolstrup, E. 2005. Palsa Development and Associated Vegetation in Northern Sweden. *Arctic, Antarctica, and Alpine Research* **37**: 49-60.

**Figures**

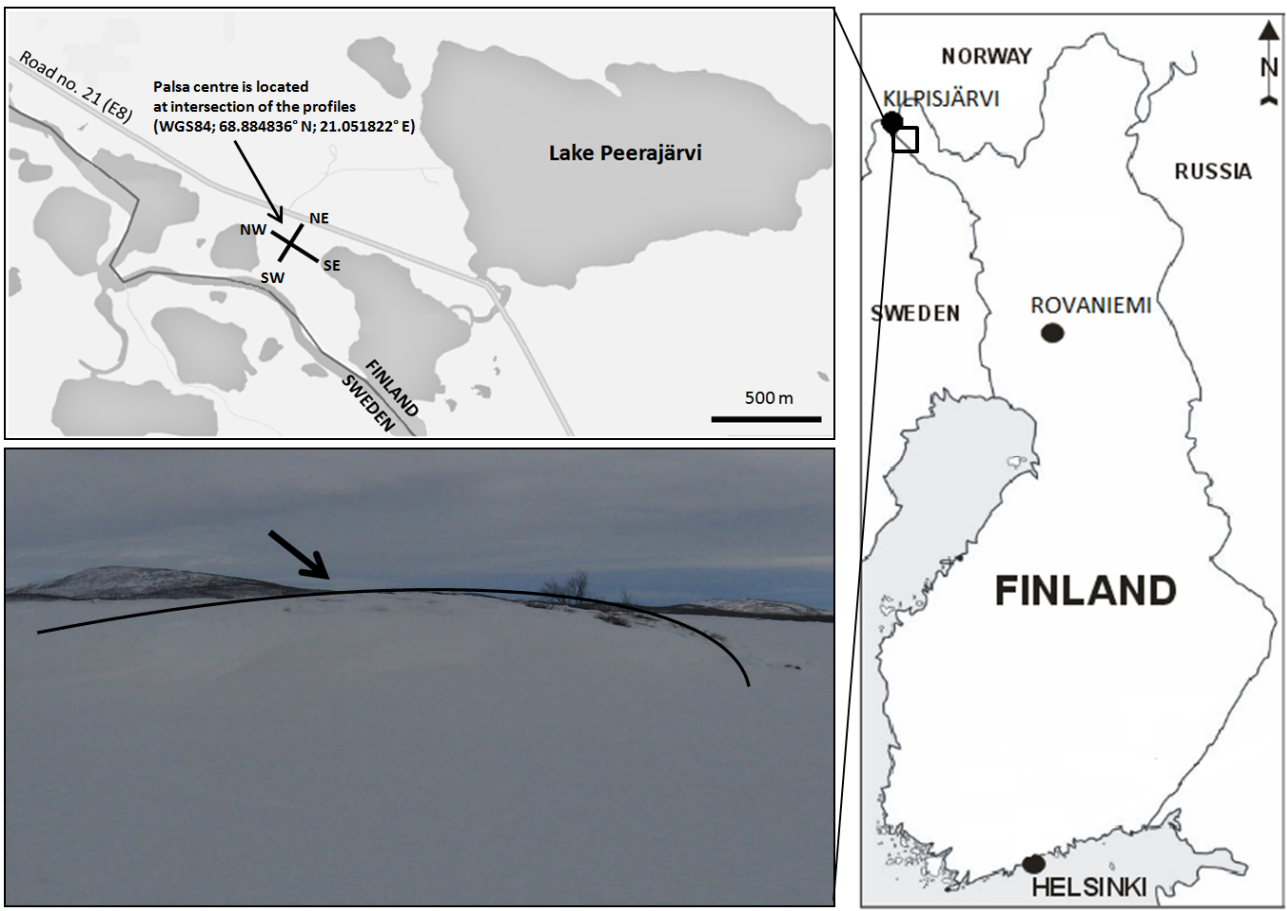

Figure 1. A photograph of the study palsa and a map of the area with the position of the profiles (black straight lines). The horizontal dimensions of the palsa are approximately 30 x 60 m and its height above the peat surface is approximately 3 m.

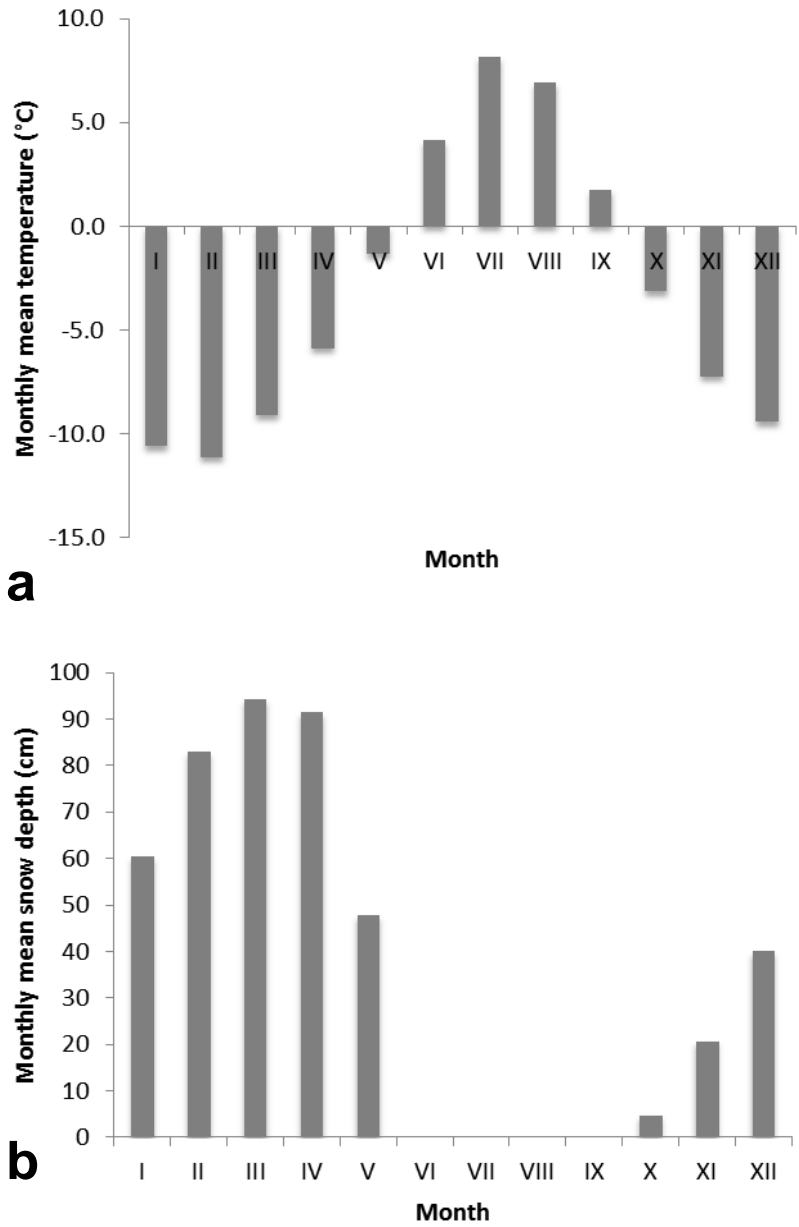

Figure 2. The monthly mean temperature (a) and snow depth (b) in Kilpisjärvi. Months are shown with Roman numerals.

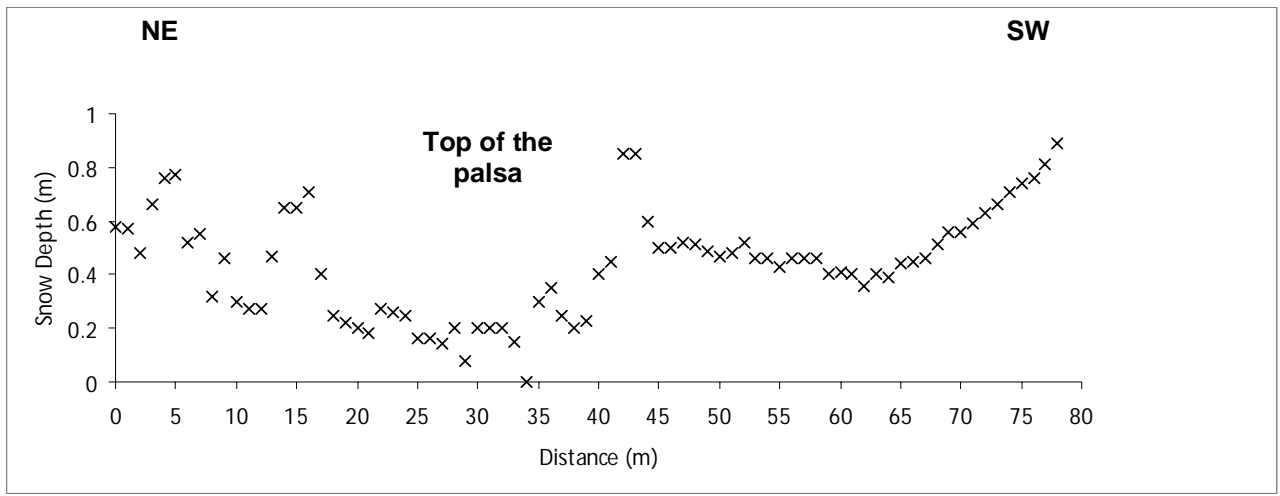

Figure 3. The snow depth in NE-SW direction as measured using a snow probe. The location of the top of the palsa is indicated.

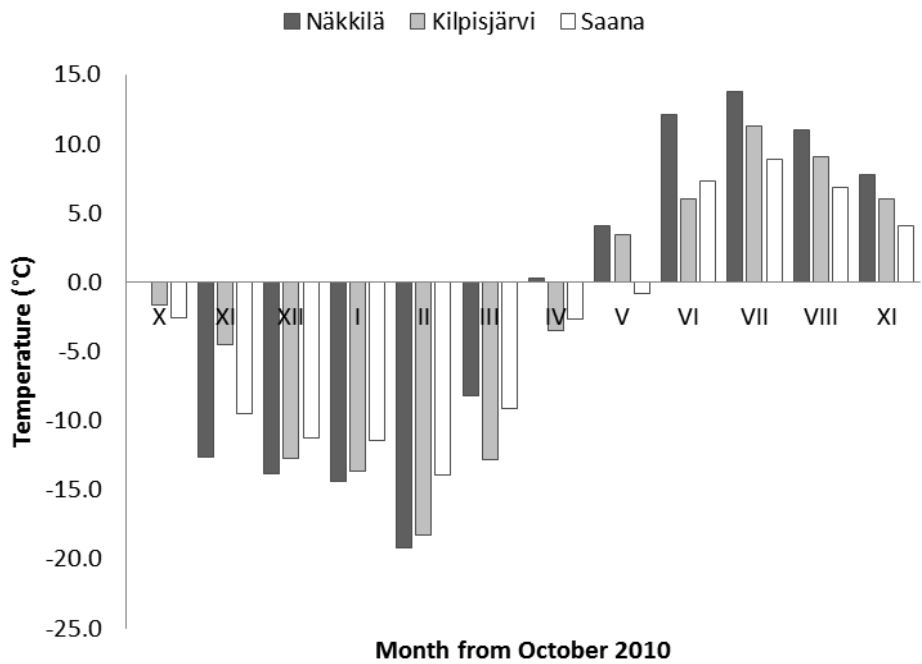

Figure 4. Monthly mean air temperature for the period October 2010 – September 2011 in Enontekiö Näkkilä, Kilpisjärvi and Saana weather stations. Months are shown with Roman numerals.

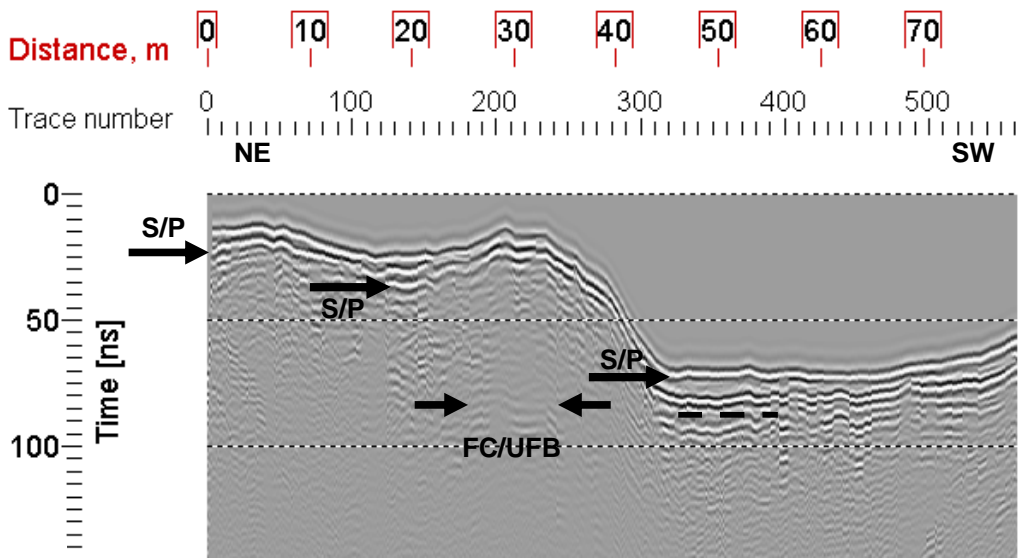

Figure 5. The winter radargram through the palsa in a NE-SW direction. The reflectors, interpreted as the snow/peat interface (S/P) and the frozen core/unfrozen base interface (FC/UFB) are indicated with arrows. The reflector of double two-way travel time and reversed polarity, marked with a thick dashed line, is a surface multiple reflection.

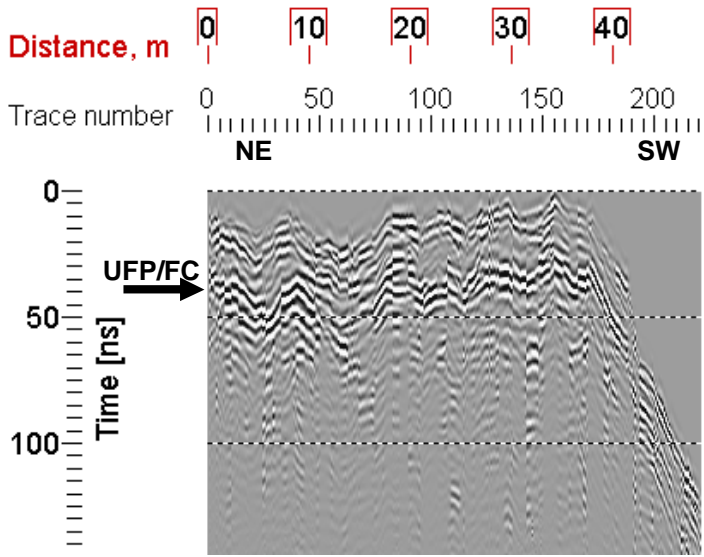

Figure 6. The summer radargram through the palsa in a NE-SW direction. The reflector interpreted as unfrozen peat/frozen core (UFP/FC) is indicated with an arrow.

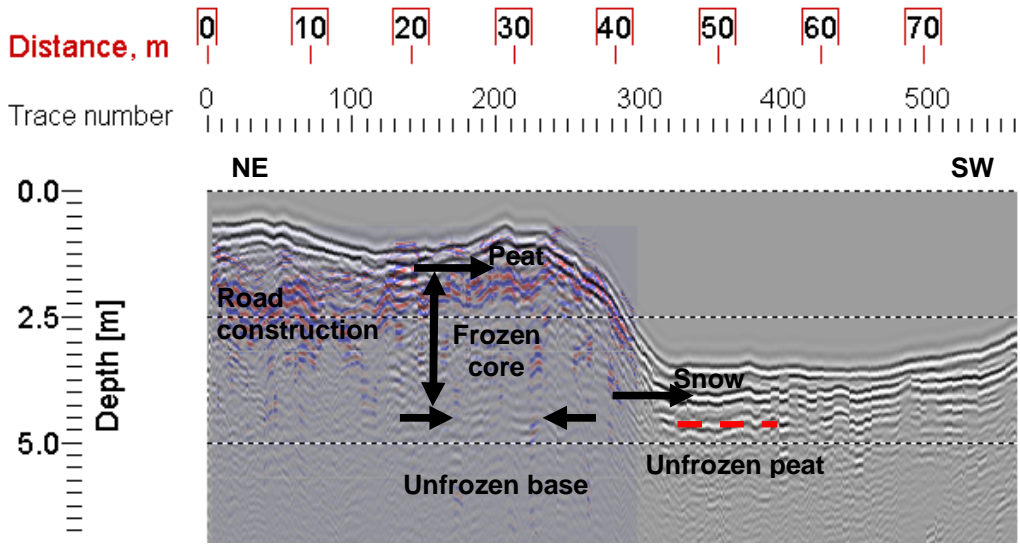

Figure 7. Combined summer and winter radar profile (after time to depth conversion) in the NE-SW direction through the palsa. The summer radargram presented in red and blue is superimposed over the winter radargram presented in black and white. In this combined profile all the main units of the palsa and their boundaries are marked. The left part of the profile is affected by construction of the adjacent road. The reflector of double two-way travel time and reversed polarity, marked with a thick red dashed line, is a surface multiple reflection.

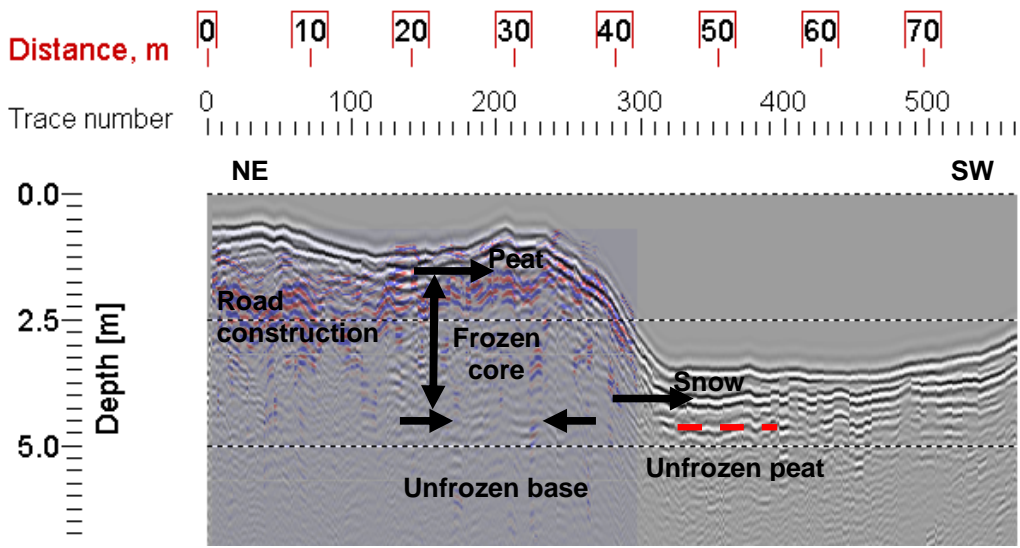

Figure 8. The winter 2013 radargram of the palsa in the NW-SE direction, perpendicular to those in Figures 5 to 7. The reflectors, interpreted as the snow/peat interface (S/P) and the frozen core/unfrozen base interface (FC/UFB) are indicated with arrows. The lateral extent of the frozen core in the perpendicular direction is at least 30 m.

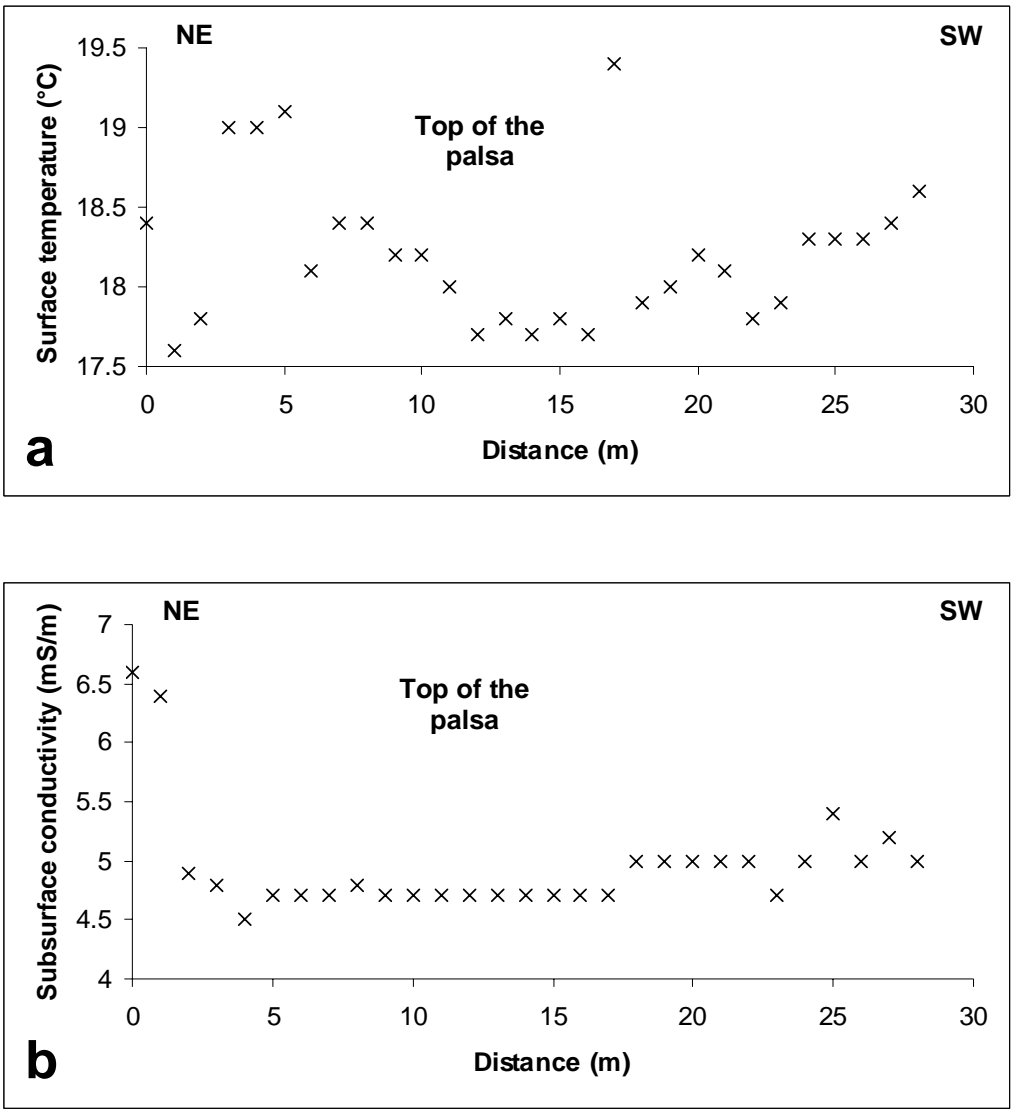

Figure 9. The summer (August 2011) surface infrared temperature (a) and ground conductivity (b) profile over the palsa in a NE-SW direction. The position of the top of the palsa is indicated. Note: The starting point of the profile is offset compared to that in Figures 4-7.